\documentclass[letterpaper,10pt,conference]{IEEEtran}
\IEEEoverridecommandlockouts

\usepackage{cite}
\usepackage{amsmath,amssymb,amsfonts}
\usepackage{algorithmic}
\usepackage{graphicx}
\usepackage{textcomp}
\usepackage{xcolor}
\usepackage{booktabs}
\usepackage{multirow}
\usepackage{circledtext}
\usepackage{tikz}
\newcommand{\CircledText}[2][white]{%
  \tikz[baseline=(char.base)]{
    \node[shape=circle, draw, fill=black, text=#1, inner sep=0.1pt] (char) {#2};
  }%
}
\def\BibTeX{{\rm B\kern-.05em{\sc i\kern-.025em b}\kern-.08em
    T\kern-.1667em\lower.7ex\hbox{E}\kern-.125emX}}
\begin{document}

\title{TherMapNet: Attention-Guided Runtime Full-Chip Thermal Map Prediction from Performance Metrics
\thanks{*Corresponding author: Lin Jiang.}
}

\author{\IEEEauthorblockN{Qin Gu\textsuperscript{1},
Chaofang Ma\textsuperscript{2},
Mingyu Yang\textsuperscript{1},
Yipu Zhang\textsuperscript{2},
Jiliang Zhang\textsuperscript{1},
Wei Zhang\textsuperscript{2},
and Lin Jiang\textsuperscript{1,*}}
\IEEEauthorblockA{\textsuperscript{1}College of Information Science and Engineering,
Northeastern University, Shenyang, China \\
\textsuperscript{2}Department of Electronic and Computer Engineering,
The Hong Kong University of Science and Technology, Hong Kong \\
E-mail: 18380126229@163.com,
cmaaw@connect.ust.hk,
yangmingyu@mails.neu.edu.cn, \\
yzhangqq@connect.ust.hk,
zhangjiliang1@mail.neu.edu.cn,
eeweiz@ust.hk,
jianglin1@neu.edu.cn}
}

\maketitle

\begin{abstract}
As the power density of high-performance chips continues to increase and air-cooling technologies approach their theoretical limits, runtime thermal management has become increasingly critical for maintaining chip performance and prolonging chip lifespan. This, however, relies heavily on fast and accurate full-chip thermal simulators capable of offering thermal maps in runtime. In this work, TherMapNet, an attention-guided thermal simulator, is proposed to directly predict full-chip thermal maps from performance metrics with high accuracy and low latency, unlike conventional thermal simulators that incur additional overhead to estimate power traces from performance metrics. TherMapNet employs a Transformer encoder to capture the temporal evolution of thermal maps, followed by a convolutional neural network (CNN) for fine-grained spatial feature extraction. In the Transformer encoder, the temporal sequence of each performance metric is treated as an individual token. This representation enables TherMapNet to more effectively capture temporal dependencies and dynamic workload variations, thereby improving dynamic thermal estimation capability. For the CNN module, a dual-branch channel-spatial attention convolution module (DACM) and triplet loss are designated to significantly enhance the spatial feature learning efficiency, ultimately improving the accuracy of TherMapNet. To demonstrate the effectiveness of TherMapNet, it is applied to thermal map prediction for a multi-core CPU (AMD Ryzen 7 4800U) and a many-core GPU (NVIDIA GeForce RTX 4060). Experimental results show that TherMapNet significantly outperforms state-of-the-art thermal simulators, achieving a root mean square error below 0.26°C while requiring less than 2.4 ms inference time on an NVIDIA GeForce RTX 3090 GPU. These results reveal the potential of TherMapNet for enabling high-quality runtime thermal management in modern multi-core chips.
\end{abstract}


\section{Introduction}
The ever-growing computing demands of AI, high-performance computing, and big data have driven the continuous advancement of chip technology nodes, operating frequency, and voltage in recent decades~\cite{chip_ai}. This advancement, however, poses significant thermal challenges on modern multi-core chips, where high temperature not only degrades performance, impairs reliability, but also raises power consumption~\cite{temperature_in_chips2,sultan2019survey}. To address the thermal issues, various thermal management techniques have been developed, for instance dynamic voltage and frequency scaling~\cite{tan2024thermal}. Such effective thermal management techniques strongly drive the need of a fast and accurate thermal estimator to provide real-time temperature profiles of chips~\cite{sultan2019survey}.

 \begin{figure}[htbp]
    \centering
    \includegraphics[width=\linewidth]{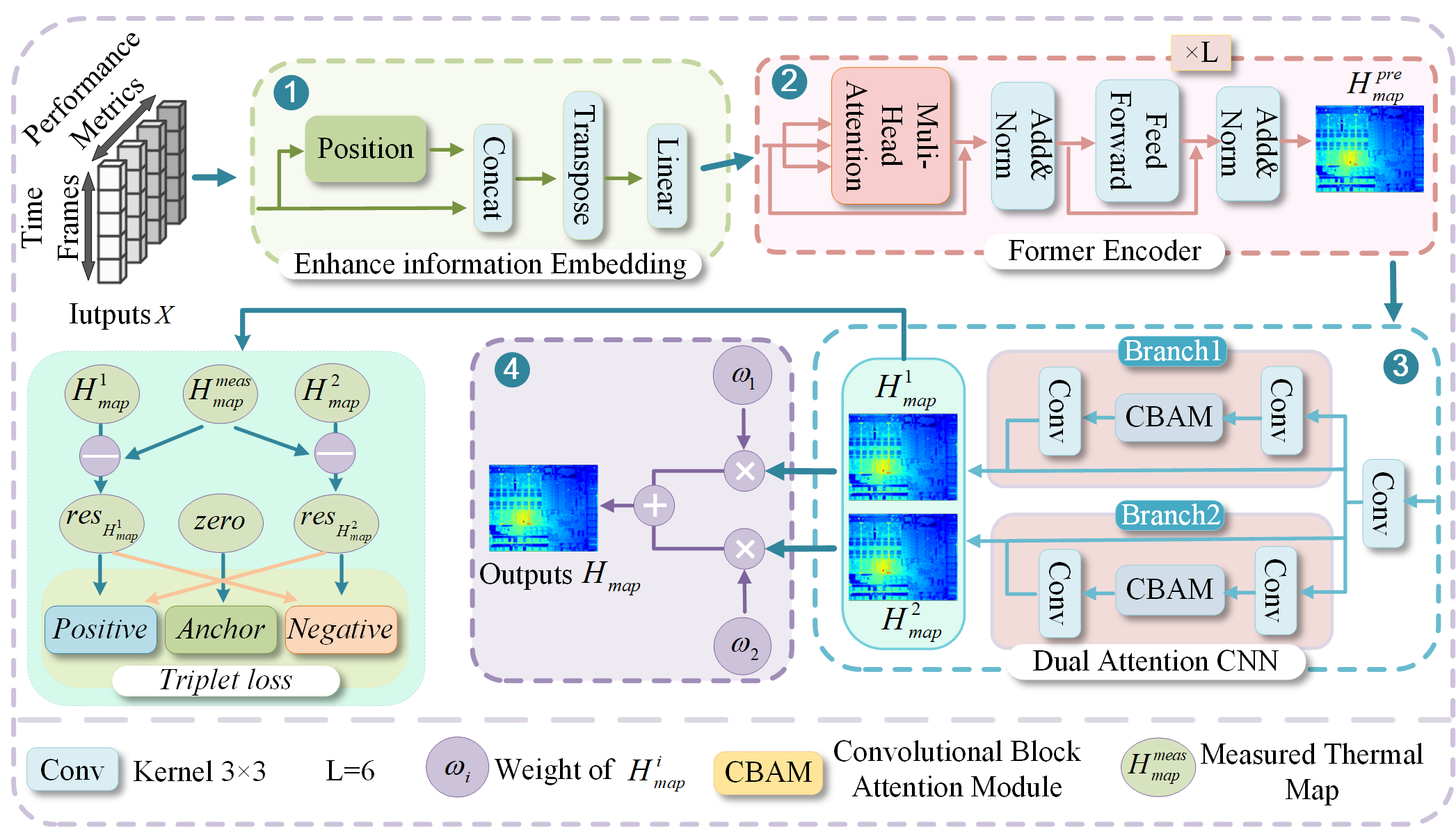}
    \caption{The Complete Architecture of TherMapNet.}
    \label{fig:architecture}
\end{figure}

{Over the past decades, a variety of thermal simulators have been developed for chip thermal analysis, typically emphasizing either accuracy or efficiency. These include direct numerical simulations based on the finite element method, the Green's function method, and thermal circuit models. None of these, however, can simultaneously deliver both high accuracy and high efficiency. For example, thermal circuit models achieve high efficiency at the expense of accuracy. Recently, the proper orthogonal decomposition (POD) method has been introduced for chip thermal estimation, demonstrating both high accuracy and efficiency~\cite{jiang2023podtherm}. However, the POD-based approaches necessitate a prior power estimation process using chip power simulators such as McPAT~\cite{jiang2024ensemble}, which introduces additional computational overhead.

With the rapid advancement of deep learning techniques, especially neural networks (NNs), NN-based thermal simulators have emerged as promising alternatives for chip thermal estimation. Notable examples include the long-short term memory (LSTM)-based RealMaps~\cite{sadiqbatcha2021realmaps}, the generative adversarial network (GAN)-based ThermalGAN~\cite{jin2020ThermalGAN}, the transformer-based GPUThermalMap~\cite{lu2025GPUThermalMap} and ThermalTransformer~\cite{lu2023ThermalTransformer}, as well as the operator-learning-based DeepOHeat~\cite{liu2023deepoheat}. Despite these developments, existing NN-based approaches present some limitations in terms of accuracy and/or efficiency for dynamic thermal management of modern chips. For instance, RealMaps suffers from limited accuracy, while ThermalGAN is more suited for steady-state thermal estimation due to its reduced capability in handling dynamic data.

To deal with these limitations, this work proposes TherMapNet, a transformer–convolutional neural network (CNN) cooperative thermal simulator that delivers accurate and efficient real-time chip temperature profiles directly from chip performance metrics collected from on-chip performance monitors, including Intel’s Performance Counter Monitor (PCM)~\cite{pcm}. The complete architecture of TherMapNet is presented in Fig.~\ref{fig:architecture}. As illustrated, unlike ThermalTransformer, which takes all performance metrics at a single instant as one token, TherMapNet represents the time series of each individual metric across multiple steps as a token (Steps \CircledText[white]{1} and \CircledText[white]{2} in Fig.~\ref{fig:architecture}). This temporal representation enables TherMapNet to capture temporal correlations and dynamic variations more effectively, thereby enhancing its capability for dynamic thermal estimation. In addition, a CNN layer is incorporated in TherMapNet to enhance the spatial information of chip temperature profiles (Steps \CircledText[white]{3} and \CircledText[white]{4} in Fig.~\ref{fig:architecture}). Leveraging CNN's strong spatial feature extraction, TherMapNet achieves significantly higher accuracy than transformer-only simulators.

The main contributions of this work are as follows:}

\begin{itemize}
\item  \textbf{TherMapNet}, a transformer–CNN cooperative thermal simulator, is proposed, directly offering chip temperature from performance metrics. It leverages a transformer module to capture temporal correlations and dynamic variations of performance metrics, and a CNN module to enhance the spatial features of temperature distributions.  
\item TherMapNet represents the time series of each individual metric as a token, enabling the transformer module to capture temporal variations more effectively. In addition, two types of positional encoding are introduced to better model the influence of time-varying performance indicators on thermal maps.  
\item A dual-branch channel–spatial attention convolution module (DACM) is designated to effectively capture spatial features of thermal distributions through differentiated feature extraction and fusion. Each branch integrates residual connections with channel–spatial attention mechanisms to extract rich thermal map features, while their outputs are fused using adaptive feature weighting. Moreover, a triplet loss is proposed to further improve the accuracy of thermal map reconstruction.  
\end{itemize}

\section{Related Works}
Finite element method (FEM)-based direct numerical simulations (DNSs) are adopted for chip thermal estimations owing to their high accuracy. However, achieving this accuracy requires extremely fine meshes, resulting in a large number of degrees of freedom and thereby making DNS computationally prohibitive for real-time chip thermal estimation. Consequently, DNS is typically used as a validation reference rather than a practical solution. For example, Chakraborty et al.~\cite{chakraborty2024treafet} employed DNS to validate a temperature-aware task scheduling approach for multi-core chips, ensuring that thermal constraints were satisfied in scheduling.

To improve the efficiency of thermal estimations, thermal circuit models and the Green’s function method have been proposed, but these approaches trade off accuracy for computational speed. Thermal circuit models simplify the chip by lumping functional blocks into single thermal nodes, which however leads to substantial errors in modeling heat transfer between nodes; Shang et al.~\cite{Shang2004} reported errors exceeding 200\% for certain floorplans compared to FEM results. The Green’s function method estimates temperature profiles by convolving a precomputed Green’s function with the chip power map, assuming an infinite chip and ignoring realistic boundary conditions, which results in significant edge and corner errors~\cite{sultan2019survey}. While some efforts have been made to handle the edge and corner errors, they are typically limited to adiabatic boundary conditions~\cite{ziabari2014power}.

Proper orthogonal decomposition (POD)-based methods have been applied at both chip and device levels~\cite{jiang2023podtherm,jia2022methodology}. These methods project thermal estimation problems from the computationally expensive physical space onto a low-dimensional function space, achieving speedups of up to three orders of magnitude. By incorporating the heat transfer equation via Galerkin projection, POD-based approaches are physically guided and can maintain least-squares errors below 1\% both within and beyond the training range~\cite{jiang2024ensemble}. However, as to thermal management of chips, a separate dynamic power estimator is required to provide power maps for POD-based approaches, introducing additional computational overhead~\cite{jiang2023podtherm}.

Recently, several NN-based thermal simulators enabled by deep learning techniques have been proposed to directly predict chip temperature profiles from real-time utilization and monitoring performance data, such as core frequency, voltage, and performance counters~\cite{pcm,uProf}. RealMaps~\cite{sadiqbatcha2021realmaps} uses temporal-aware LSTMs and shows promise for real-time thermal management, but LSTMs struggle with long-range temporal dependencies, leading to cumulative errors in transient thermal simulations. ThermalGAN~\cite{jin2020ThermalGAN} leverages generative adversarial networks for improved accuracy, yet GANs are designed for spatial rather than temporal modeling, limiting their effectiveness in dynamic scenarios. Transformer-based simulators, such as ThermalTransformer~\cite{lu2023ThermalTransformer} and GPUThermalMap~\cite{lu2025GPUThermalMap}, address temporal challenges using attention mechanisms to capture long-range temporal dependencies. Nevertheless, their spatial feature extraction is limited, often producing blurred thermal gradients.

\section{TherMapNet Framework}
\subsection{Overview}

 In this work, TherMapNet, a transformer–CNN cooperative thermal simulator, is proposed to directly provide thermal estimation of multi/many-core chips from their performance metrics with high accuracy and efficiency. The overall framework of TherMapNet is illustrated in Fig.~\ref{fig:architecture}. Specifically, the collected dynamic performance metrics are organized as a matrix, denoted by $\boldsymbol{X} = \{\boldsymbol{x}_1, \dots, \boldsymbol{x}_T\} \in \mathbb{R}^{T \times N}$, where $\boldsymbol{x}_{i}\in \mathbb{R}^{N}$  represents the performance metric vector at the $i$-th time frame, $T$ is the total number of time frames, and $N$ is the number of performance metrics. The performance metric matrix is first processed by the enhanced Transformer architecture of TherMapNet to capture temporal dependencies, generating a preliminary thermal map, $H_{\text{map}}^{\text{pre}}$. This preliminary thermal map is subsequently refined by a dual-branch CNN to extract spatial features. To improve the learning effectiveness of spatial features, a triplet-loss constraint is introduced to encourage the two branches to learn complementary and non-overlapping feature representations, producing $H_{\text{map}}^{1}$ and $H_{\text{map}}^{2}$, respectively. Finally, these two thermal maps are adaptively fused to generate the final thermal map, $H_{\text{map}}$, of the target chip.

\subsection{Absolute and Relative Positional Channels Enhanced Transformer}
The accuracy of the preliminary thermal map largely depends on the capability of Transformer encoder to capture temporal dependencies embedded in the collected performance metrics. To strengthen this capability, TherMapNet incorporates both absolute and relative positional information into the Transformer encoder. Specifically, an absolute positional channel is introduced at the input stage to explicitly encode the temporal order of performance metrics, while a relative positional channel is integrated into the self-attention mechanism to model pairwise temporal relationships between tokens. The combination of these two positional encoding schemes enables TherMapNet to more effectively capture both global and local temporal dependencies.
The two positional encoding mechanisms are detailed in the remainder of this subsection.

\subsubsection{Absolute Positional Channel}
Existing Transformer-based thermal simulators, such as ThermTransformer~\cite{lu2023ThermalTransformer}, concatenate all performance metrics collected at a given time frame and treat the resulting vector as a single token. Although this formulation preserves the temporal order of the sequence, it weakens the model's sensitivity to the temporal evolution of individual performance metrics. Since different performance metrics characterize distinct physical activities within a chip, their temporal behaviors may contribute differently to thermal dynamics. Consequently, representing all metrics as a single token limits the ability of the Transformer to capture metric-specific temporal patterns. 

\begin{figure}[tbp]
    \centering
    \includegraphics[width=0.95\linewidth]{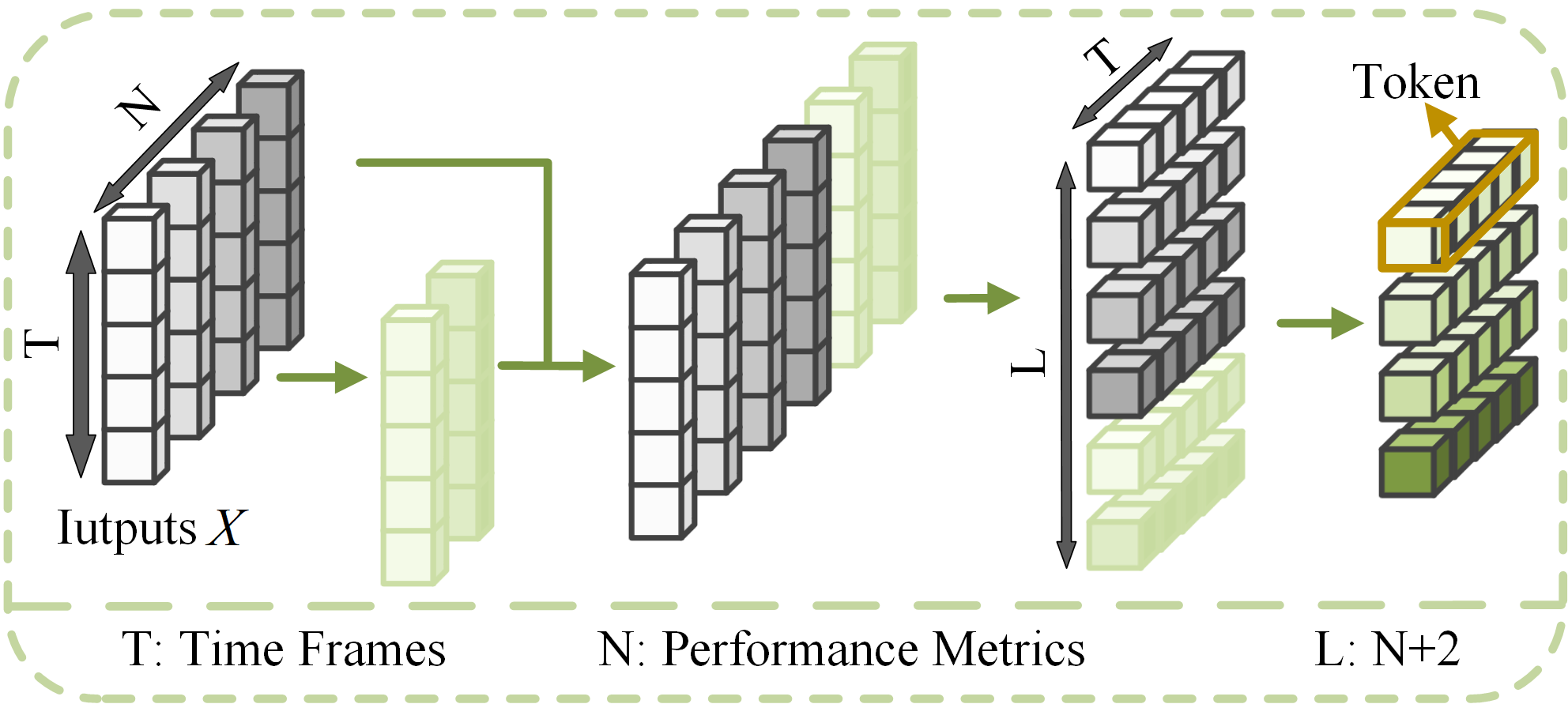} 
    \caption{Enhance information embedding.}
    \label{fig:embed}
\end{figure}

To address this limitation, TherMapNet adopts a metric-centric tokenization strategy. As illustrated in Fig.~\ref{fig:embed}, the time series associated with each performance metric across multiple time frames is treated as an individual token. This design enables the Transformer encoder to directly model the temporal evolution of each performance metric, thereby improving its capability to capture fine-grained temporal dynamics and enhancing the fidelity of thermal map estimation. 
\begin{align}
    \text{pos}(\boldsymbol{X}) = \left(\sin\frac{2\pi t}{P}, \cos\frac{2\pi t}{P}\right), \quad \text{pos}(\boldsymbol{X})\in \mathbb{R}^{T \times 2},
\end{align}
where $\text{pos}(\boldsymbol{X})$ denotes the encoded positional information and $P$ is the cycle length determined by the sampling period used for collecting performance metrics. The sinusoidal formulation constrains the positional encoding values to the range $[-1,1]$, matching the scale of the normalized input features. As a result, the encoded positions provide explicit temporal order information without introducing numerical imbalance among feature dimensions. As illustrated in Fig.~\ref{fig:embed}, the encoded absolute positional information is subsequently concatenated with collected performance metrics  
\begin{align}
    \boldsymbol{X}' &= [\boldsymbol{X} ~||~ \text{pos}(\boldsymbol{X})], \quad \boldsymbol{X}' \in \mathbb{R}^{T \times L},
\end{align}
where $\boldsymbol{X}'$ is the concatenated matrix, $||$ represents the concatenation operation, and $L = N+2$. The concatenated matrix $\boldsymbol{X}'$ is then transposed, and the time series of each individual metric is treated as a dedicated token
\begin{align}
    \boldsymbol{X}_{\text{emb}} &= \text{Linear}~\!\big({\boldsymbol{X}'}^\top \big), \quad \boldsymbol{X}_{\text{emb}} \in \mathbb{R}^{L \times d_x},
\end{align}
where $\boldsymbol{X}_{\text{emb}}$ denotes the token embedding matrix, $\text{Linear}(\cdot)$ represents a linear projection, ${\boldsymbol{X}'}^\top$ is the transpose of $\boldsymbol{X}'$, and $d_x$ is the embedding dimension.

\subsubsection{Relative Positional Channel}
Although the absolute positional channel explicitly embeds the temporal order of performance metrics, it only provides information about the absolute location of each time frame in the observation window. It does not directly model the relative temporal distance between different tokens, which is often crucial for capturing temporal dependencies. For example, two tokens that are temporally adjacent may exhibit stronger correlations than those separated by a larger temporal interval, regardless of their absolute positions. Therefore, relying solely on absolute positional information may limit the Transformer's ability to accurately characterize temporal relationships among performance metrics. To handle this limitation, a relative positional channel is further incorporated into the self-attention mechanism of TherMapNet following~\cite{shaw2018self}. Combined with the absolute positional channel introduced at the input stage, the relative positional channel enables the Transformer encoder to capture both absolute temporal order and pairwise temporal relationships, thereby strengthening its capability to model temporal dependencies.

\begin{figure}[tbp]
    \centering
    \includegraphics[width=0.95\linewidth]{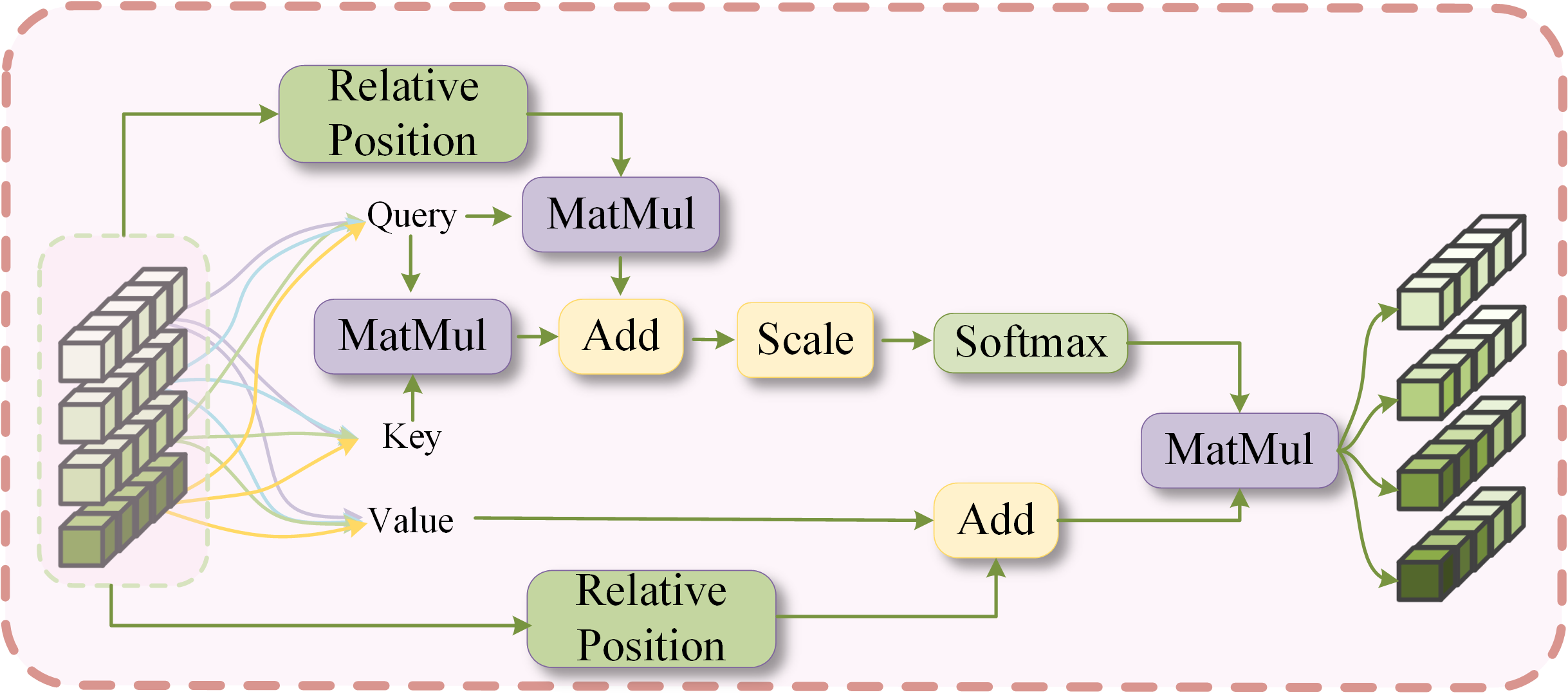} 
    \caption{Self attention with relative position.}
    \label{fig:attention}
\end{figure}

As illustrated in Figs.~\ref{fig:architecture} and~\ref{fig:attention}, the Transformer encoder of TherMapNet adopts the standard multi-head self-attention architecture. Let $\boldsymbol{X}_{\text{emb}}=\{\boldsymbol{x}_1,\ldots,\boldsymbol{x}_L\}$ be the input sequence and the corresponding output sequence be denoted by $\boldsymbol{Z}=\{\boldsymbol{z}_1,\ldots,\boldsymbol{z}_L\}$, where $\boldsymbol{z}_i\in\mathbb{R}^{d_z}$. Similar to the standard self-attention mechanism, each input vector is projected into a query $\boldsymbol{q}_i$, a key $\boldsymbol{k}_i$, and a value $\boldsymbol{v}_i$:
\begin{equation}
\boldsymbol{q}_i = \boldsymbol{x}_i \boldsymbol{W}^{Q}, \quad \boldsymbol{k}_i = \boldsymbol{x}_i \boldsymbol{W}^{K}, \quad \boldsymbol{v}_i = \boldsymbol{x}_i \boldsymbol{W}^{V},
\label{eq:qkv}
\end{equation}
where $\boldsymbol{W}^{Q}$, $\boldsymbol{W}^{K}$, and $\boldsymbol{W}^{V} \in \mathbb{R}^{d_x \times d_z}$ are learnable weight matrices. To explicitly encode pairwise temporal relationships, the relative positional channel introduces learnable relative position encodings into both the key and value representations. For each pair of tokens $(\boldsymbol{x}_i,\boldsymbol{x}_j)$, the corresponding relative key and value embeddings are defined as

\begin{equation}
\left\{
\begin{aligned}
\boldsymbol{a}_{ij}^{K} &= \boldsymbol{w}_{clip(j - i, ~k)}^{K},\\
 \boldsymbol{a}_{ij}^{V} &= \boldsymbol{w}_{clip(j - i, ~k)}^{V},\\
 \text{clip}[(j -i ), k] &= \max(-k, \min(k, (j -i ))),
\end{aligned}
\right.
\label{eq:relative representations}
\end{equation}
where $\boldsymbol{a}_{ij}^{K}$ and $\boldsymbol{a}_{ij}^{V}$ denote the relative key and value embeddings, respectively. The clipping operation limits the relative distance to the range $[-k,k]$, preventing excessive parameter growth while preserving local temporal relationships. Once the relative key and value embeddings are determined, the compatibility score between $\boldsymbol{x}_i$ and $\boldsymbol{x}_j$ is given by
\begin{equation}
    e_{ij} = \frac{\boldsymbol{q}_i \left( \boldsymbol{k}_j + \boldsymbol{a}_{ij}^{K} \right)^{\mathrm{T}}}{\sqrt{d_z}}.
\label{eq:attn_score}
\end{equation}
The attention weight $\alpha_{ij}$ is then computed as
\begin{equation}
    \alpha_{ij} = \frac{\exp(e_{ij})}{\sum_{k=1}^{L} \exp(e_{ik})}.
\label{eq:attn_weights}
\end{equation}
The final output of $\boldsymbol{z}_i$ is obtained by aggregating the value vectors and the corresponding relative position encodings
\begin{equation}
    \boldsymbol{z}_i = \sum_{j=1}^{L} \alpha_{ij} \left( \boldsymbol{v}_j + \boldsymbol{a}_{ij}^{V} \right).
\label{eq:attn_output}
\end{equation}

\subsection{Dual Branch Channel Spatial Attention Convolution Module (DACM)}
The Transformer encoder equipped with absolute and relative positional channels primarily focuses on modeling temporal dependencies among performance metrics and generates a preliminary thermal map, $H_{\text{map}}^{\text{pre}}$ (Step \CircledText[white]{2} in Fig.~\ref{fig:architecture}). Although $H_{\text{map}}^{\text{pre}}$ captures the global thermal distribution, it is less effective in recovering fine-grained spatial details, such as local temperature gradients and hotspot boundaries. To further enhance spatial feature extraction, a Dual Branch Channel Spatial Attention Convolution Module (DACM) is introduced in TherMapNet.

\begin{figure}[tbp]
    \centering
    \includegraphics[width=0.85\linewidth]{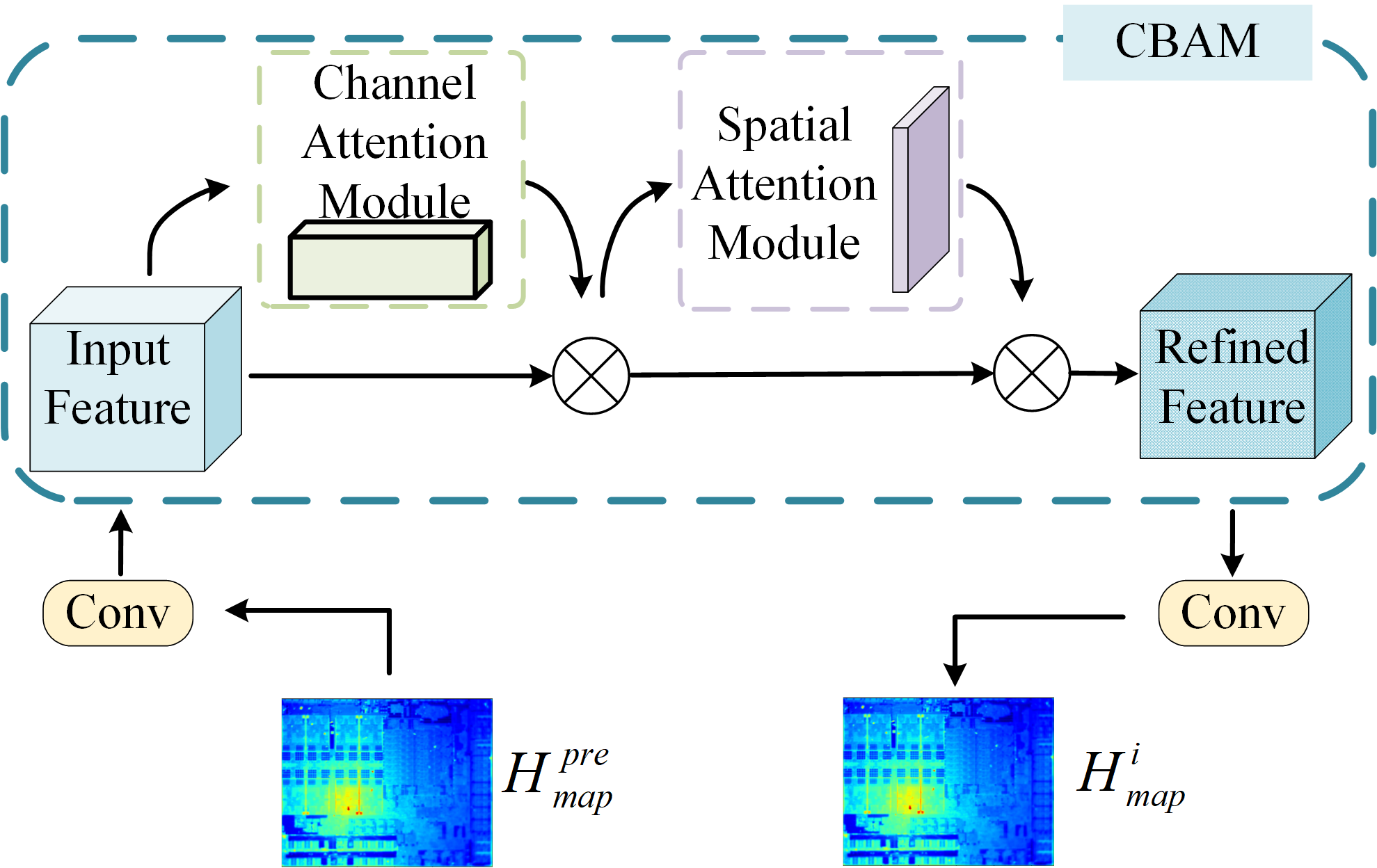} 
    \caption{One branch of the DACM.}
    \label{fig:DACM}
\end{figure}

As illustrated in Step \CircledText[white]{3} of Fig.~\ref{fig:architecture}, the preliminary thermal map $H_{\text{map}}^{\text{pre}}$ is first processed by several convolutional layers to extract shallow spatial features. The resulting feature maps are then propagated through two parallel CNN branches to learn complementary high-level spatial representations. To improve the discriminative capability of each branch, a convolutional block attention module (CBAM)~\cite{woo2018cbam} is incorporated into both branches, as shown in Fig.~\ref{fig:architecture}. As illustrated in Fig.~\ref{fig:DACM}, the CBAM sequentially applies channel and spatial attention mechanisms to emphasize thermally relevant regions while suppressing redundant information, thereby improving spatial feature extraction.

Unlike a conventional single-branch CNN, the dual-branch architecture is designed to capture complementary spatial characteristics of the thermal map. To explicitly encourage feature diversity between the two branches, triplet-loss constraints are imposed during training, which will be detailed in the next subsection. As a result, the two branches generate distinct thermal representations, denoted by $H_{\text{map}}^{1}$ and $H_{\text{map}}^{2}$, respectively. The resulting feature maps of $H_{\text{map}}^{1}$ and $ H_{\text{map}}^{2} $ are fused through a spatially adaptive weighted strategy to generate the final thermal map $H_{\text{map}}$
\begin{equation}
  H_{\text{map}}(l,w) = w_1(l,w)H_{\text{map}}^{1}(l,w)\\ 
 + w_2(l,w)H_{\text{map}}^{2}(l,w),
\label{eq:map_output}
\end{equation}
where $w_1(l,w)$ and $w_2(l,w)$ are learnable location-dependent fusion weights.  To ensure stability and interpretability, the weights are normalized via a softmax function across the two branches and are subject to
 \begin{equation}
\left\{
\begin{aligned}
w_1(l,w) + w_2(l,w) = 1,\\
w_1(l,w), w_2(l,w) \geq 0.
\end{aligned}
\right.
\label{eq:constrain}
\end{equation}
This adaptive fusion mechanism allows the model to dynamically exploit the strengths of each branch at different spatial locations, leading to more accurate thermal estimation.

 \subsection{Triplet Loss}
The dual-branch architecture is intended to learn complementary spatial representations. However, without additional constraints, the two branches may converge to highly similar feature spaces, reducing the benefit of using multiple branches. To address this issue, triplet-loss constraints are introduced during training to explicitly encourage feature diversity between the two branches.

Instead of applying triplet loss directly to feature embeddings, TherMapNet constructs the triplets using the residual maps of the branch outputs with respect to the measured thermal map. Specifically, the residual maps of the two branches are defined as
\begin{equation}
\left\{
\begin{aligned}
\text{Res}_1 &= H_{\text{map}}^{\text{meas}} - H_{\text{map}}^{1},\\
\text{Res}_2 &= H_{\text{map}}^{\text{meas}} - H_{\text{map}}^{2},
\end{aligned}
\right.
\label{eq:Res_1_2}
\end{equation}
where $\text{Res}_1$ and $\text{Res}_2$ denote the residual of Branches 1 and 2, respectively, defined with respect to the measured thermal map $H_{\text{map}}^{\text{meas}}$. An ideal residual map, denoted by $\text{Res}_0$, corresponds to zero estimation error (i.e., a zero matrix). For Branch 1, $\text{Res}_0$, $\text{Res}_1$, and $\text{Res}_2$ are respectively treated as the anchor, positive, and negative samples to construct the triplet-loss function. Using un-squared L2 distance, the triplet loss of Branch 1 is thus formulated as
\begin{equation}
\begin{aligned}
L_{\text{trip}}^1 = \max(0, \|f(\text{Res}_0) - f(\text{Res}_1)\|_2 +&\alpha \\
- \|f(\text{Res}_0) -f(\text{Res}_2) \|_2&),
\end{aligned}
\label{eq:Triplet_1}
\end{equation}
where $f$ denotes the embedding function, $\alpha$ is margin, a hyperparameter, to enforce a minimum separation between positive and negative samples. With the triplet loss of $L_{trip}^1$, Branch 1 is consequently encouraged to minimize the L2 distance of its own representation from the ideal state while remaining distinct from the representation learned by Branch 2. Similarly, an analogous triplet-loss formulation is applied to Branch 2 (i.e., $L_{trip}^2$) by exchanging the roles of $\text{Res}_1$ and $\text{Res}_2$
\begin{equation}
\begin{aligned}
L_{\text{trip}}^2 = \max(0, \|f(\text{Res}_0) - f(\text{Res}_2)\|_2 +&\alpha\\
- \|f(\text{Res}_0) -f(\text{Res}_1) \|_2&).
\end{aligned}
\label{eq:Triplet_2}
\end{equation}
The total triplet loss is therefore given as 
\begin{equation}
    L_{trip}^{total} =\frac{L_{trip}^1 +L_{trip}^2}{2}.
\label{eq:total_tripel_loss}
\end{equation}

 Alongside the triplet-loss function, the root mean square error (RMSE) of $L_{\text{RMSE}}$ is employed in the training 
\begin{equation}
    L_{\text{RMSE}} = \sqrt{\mathbb{E}\left[ \| H_{\text{map}}^{\text{meas}} - H_{\text{map}} \|_F^2 \right]},
\end{equation}
where $\mathbb{E} \left[\cdot \right]$ represents the expected value over all pixels in the thermal map, $H_{\text{map}}^{\text{meas}}$ and $H_{\text{map}}$ denote the measured and estimated thermal maps, respectively, and $\|\cdot\|_F$ signifies the Frobenius norm. As a result, the overall loss function for the training of TherMapNet is given by
\begin{equation}
    L_{\text{total}} = \alpha L_{\text{RMSE}} + \beta  L_{trip}^{total},
\label{eq:total_loss}
\end{equation}
where $\alpha$ and $\beta $ are positive parameters introduced to adjust the contributions of the RMSE loss function and triplet loss function. 

\section{Demonstration and Validation}
\begin{figure}[tbp]
    \centering
    \includegraphics[width=1\linewidth]{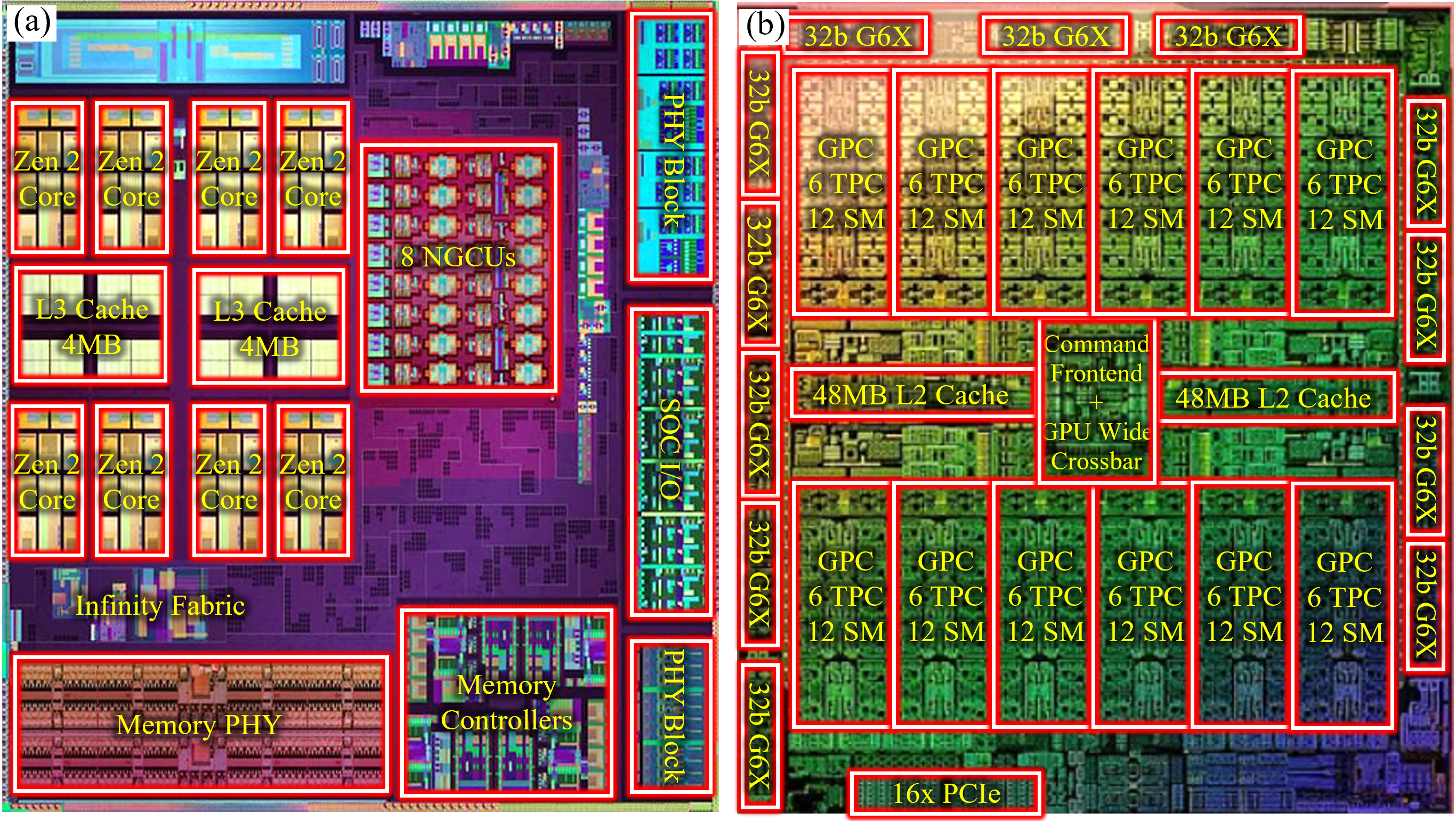}
    \caption{
    Floorplans of (a) the AMD Ryzen 7 4800U CPU (156 $mm^2$) and (b) the NVIDIA GeForce RTX 4060 GPU (159 $mm^2$). }
    \label{fig:Floorplans}
\end{figure}

To demonstrate the effectiveness of TherMapNet, it is applied to real-time thermal estimation of a multi-core CPU (i.e., AMD Ryzen 7 4800U with eight CPU cores) and a many-core GPU (i.e., NVIDIA GeForce RTX 4060 with 3,072 CUDA cores), whose floorplans are illustrated in Fig.~\ref{fig:Floorplans}. The experimental evaluation consists of three aspects. First, the prediction accuracy of TherMapNet is validated through comparisons between the predicted thermal maps and thermal infrared (IR) measurements. Then, comprehensive comparisons with state-of-the-art NN-based thermal simulators are conducted to evaluate its accuracy and inference efficiency. Finally, ablation experiments are performed to reveal the effectiveness of different architectural designs in TherMapNet. Note that all demonstration experiments were conducted on an NVIDIA GeForce RTX 3090 GPU.

\subsection{Experimental setup}
Two public datasets corresponding to the AMD Ryzen 7 4800U CPU and the NVIDIA GeForce RTX 4060 GPU~\cite{lu2024thermalMapDataset} are utilized to demonstrate TherMapNet, hereafter referred to as the AMD R7 4800U dataset and the NVIDIA RTX 4060 dataset, respectively. These datasets consist of time-series performance metrics paired with corresponding thermal maps collected from representative benchmark applications. The performance metrics are obtained using commercial monitoring tools, namely AMD uProf 4.0 and NVIDIA System Management Interface (NVIDIA SMI), while the thermal maps are captured using a thermal infrared imaging system.

For the AMD R7 4800U dataset, a total of 14,718 data samples are collected, where each sample consists of 158 performance metrics, including system-level utilization and per-core power characteristics recorded by AMD uProf 4.0, and a corresponding thermal map with a spatial resolution of $223 \times 280$ pixels. Among these samples, 11,774 are used for training TherMapNet, and the remaining samples are used for validation. For the NVIDIA RTX 4060 dataset, 16,000 data samples are collected, where each sample contains 53 performance metrics and a corresponding thermal map with a raw spatial resolution of $220 \times 300$ pixels. Due to the influence of the camera focal length, the thermal maps are cropped to a resolution of $183 \times 205$ pixels before training, with 12,800 samples used for training and 3,200 samples used for testing. Additionally, it should be noted that printed text and logos on the surface of the NVIDIA GeForce RTX 4060 GPU introduce thermal artifacts. Therefore, these regions are treated as missing thermal data during the preprocessing stage.

\subsection{Validation Against IR Measurements}
\begin{figure}[t]
    \centering
    \includegraphics[width=1\linewidth]{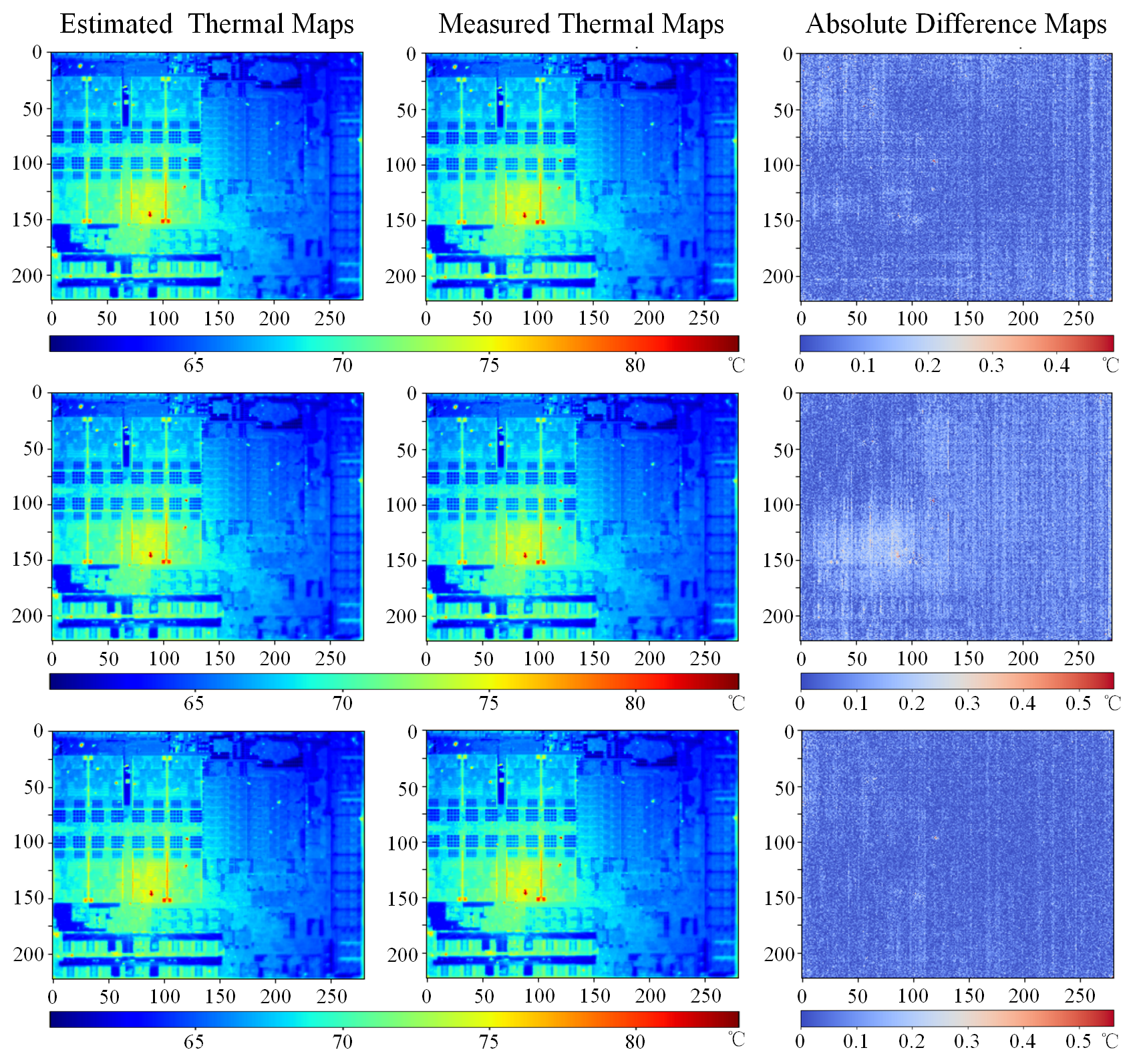}
    \caption{
    The RMSE of the estimated thermal maps for the three time steps in AMD R7 4800U are 0.071, 0.104, and 0.062 (°C). }
    \label{fig:map experiments a}
\end{figure}

\begin{figure}[t]
     \centering
    \includegraphics[width=1\linewidth]{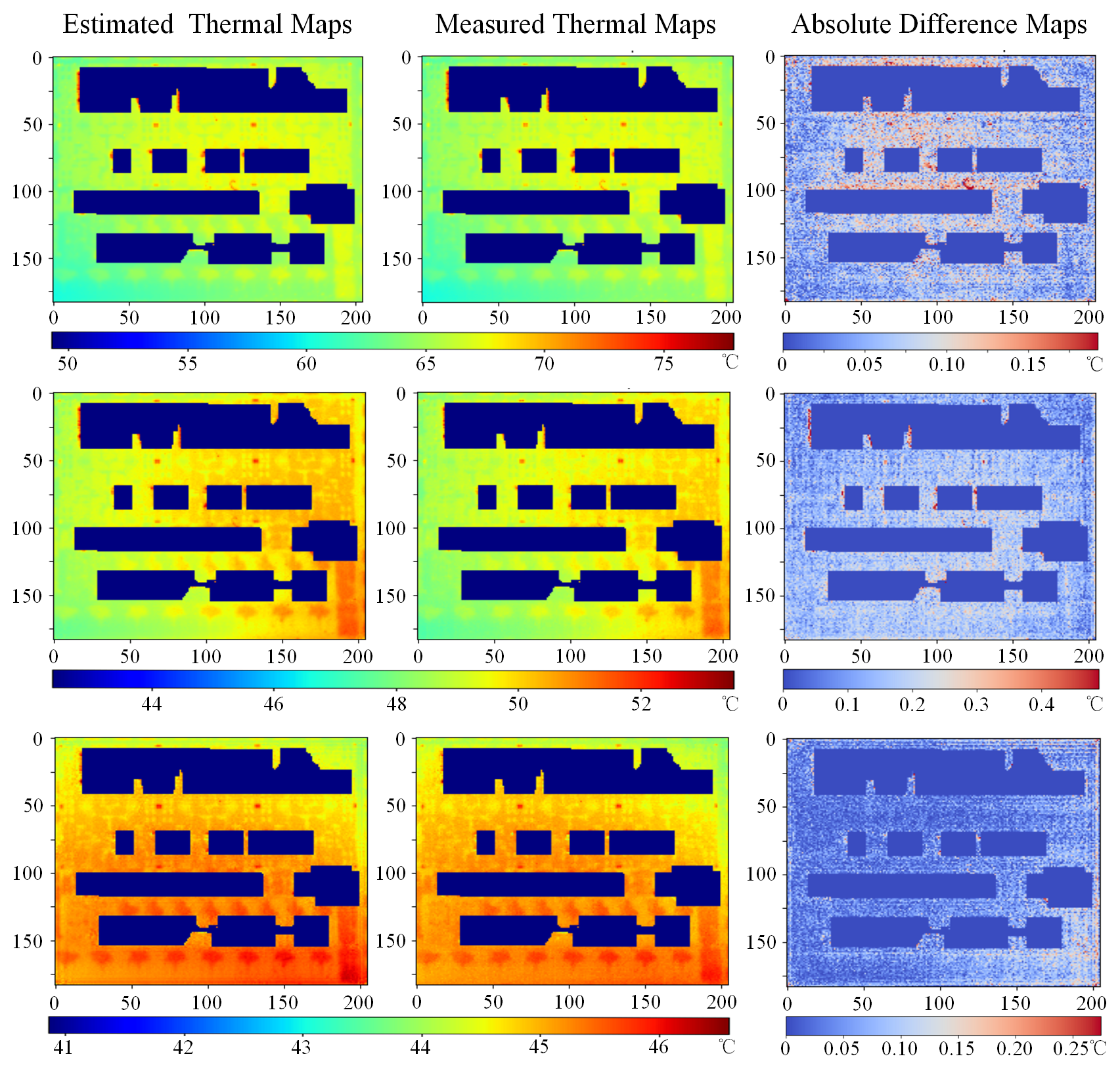}
    \caption{
    The RMSE of the estimated thermal maps for the three time steps in NVIDIA RTX 4060 are 0.027, 0.140, and 0.006 (°C). }
    \label{fig:map experiments b}
\end{figure}

The accuracy of TherMapNet is validated by comparing its predicted thermal maps of the AMD Ryzen 7 4800U CPU and the NVIDIA GeForce RTX 4060 GPU with those obtained from IR measurements, as shown in Figs.~\ref{fig:map experiments a} and~\ref{fig:map experiments b}. In Figs.~\ref{fig:map experiments a} and~\ref{fig:map experiments b}, each row corresponds to a specific time step, while the columns present the estimated thermal maps, the measured thermal maps, and the corresponding absolute error maps, respectively. The strong agreement between the thermal maps predicted by TherMapNet and those captured through IR measurements, together with the small absolute errors, highlights the capability of TherMapNet to accurately predict thermal distributions while preserving detailed spatial features.

\begin{figure}[t]
    \centering
    \includegraphics[width=0.9\linewidth]{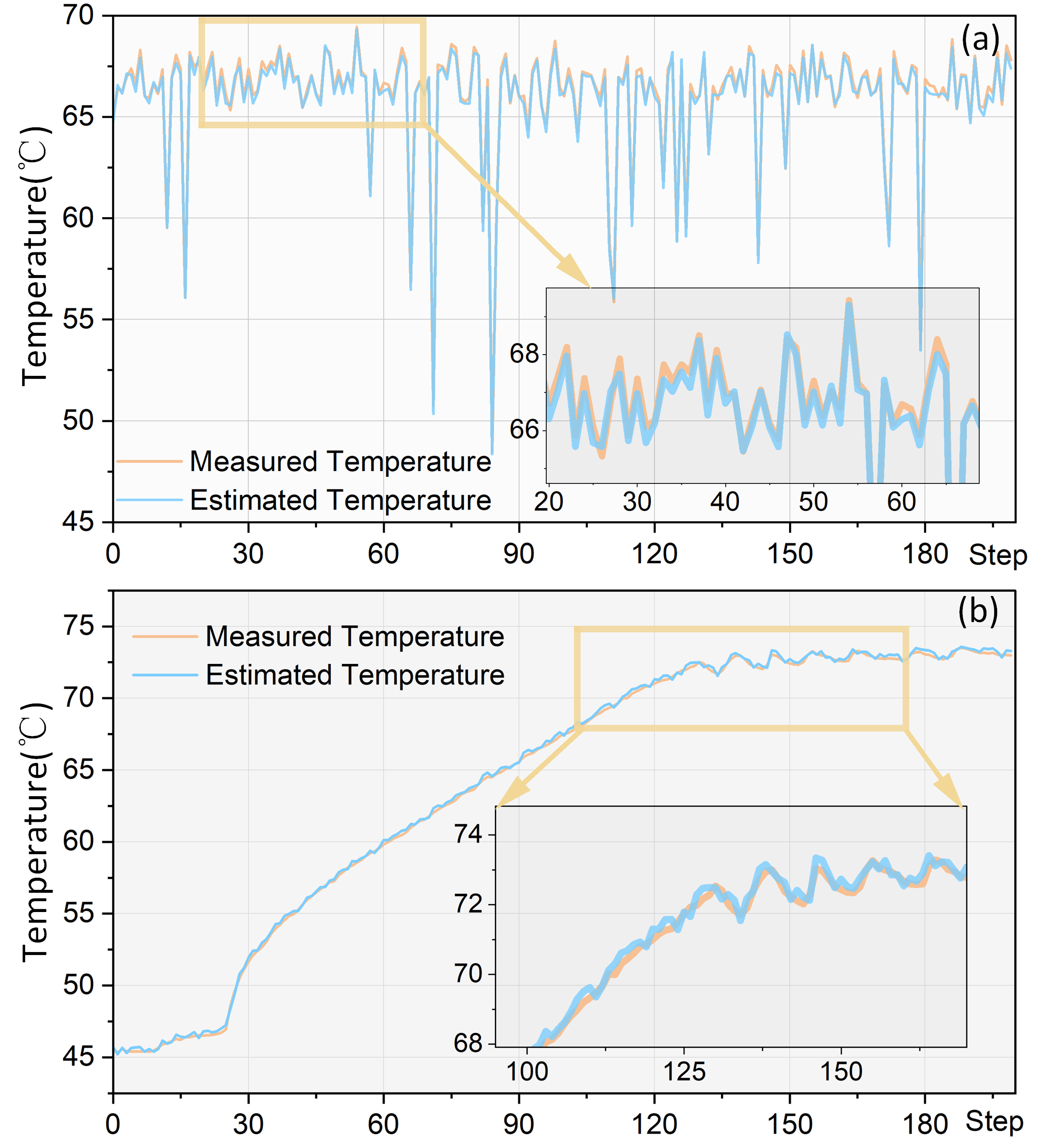}
    \caption{
    Temporal evolutions at representative pixel locations on the (a) AMD Ryzen 7 4800U and (b) NVIDIA GeForce RTX 4060.
    }
    \label{fig:estimation curve}
\end{figure}

To further evaluate the temporal prediction capability of TherMapNet, the temporal temperature evolution at selected pixel locations over 200 consecutive time steps are compared between the estimated results and IR measurements, as illustrated in Fig.~\ref{fig:estimation curve}. The results show that TherMapNet maintains consistently high estimation accuracy throughout the dynamic heating process, without noticeable error accumulation or fluctuation over time. This demonstrates its ability to reliably predict transient thermal behaviors under varying operating conditions, further confirming the robustness of TherMapNet.

\subsection{Comparisons with State-of-the-Art NN-Based Simulators}
The comparison with IR measurements confirms the capability of TherMapNet to accurately predict real-time chip thermal maps. To further benchmark its performance against existing NN-based thermal simulation approaches, state-of-the-art NN-based thermal simulators (i.e., RealMaps, ThermGAN, ThermTransformer, and GPUThermalMap) 
are also employed to perform thermal estimation for the AMD Ryzen 7 4800U CPU and the NVIDIA GeForce RTX 4060 GPU, and their estimation results are compared with those of TherMapNet. Note that, to ensure fair comparisons, all models are evaluated on the same platform with an NVIDIA GeForce RTX 3090 GPU under identical experimental settings. 

The results are presented in Table~\ref{tab:rmse_comparison}, where the root mean squared error (RMSE) are calculated across all corresponding pixels of the predicted thermal maps with respect to the measured thermal maps. It can be seen that TherMapNet achieves the best overall thermal estimation accuracy among all compared methods. For the AMD R7 4800U dataset, TherMapNet reduces the mean RMSE to 0.251$^\circ$C, outperforming ThermTransformer, ThermGAN, and RealMaps by 30.3\%, 57.9\%, and 88.5\%, respectively. It also achieves the lowest maximum RMSE of 1.429$^\circ$C and the smallest RMSE standard deviation of 0.185$^\circ$C, indicating superior robustness across different thermal maps. For the NVIDIA RTX 4060 dataset, TherMapNet similarly delivers the highest accuracy, with a mean RMSE of 0.165$^\circ$C, surpassing GPUThermalMap, ThermGAN, and RealMaps. In particular, compared with the latest GPU-specific model GPUThermalMap, TherMapNet further reduces the mean RMSE by 13.2\% and the RMSE standard deviation by 39.4\%, demonstrating its enhanced capability in capturing complex thermal distributions.

In terms of computational efficiency, RealMaps achieves the shortest inference time due to its relatively simple CNN-based architecture, but at the expense of significantly degraded accuracy. Although TherMapNet requires slightly longer inference time than some lightweight models, its inference latency remains within only a few milliseconds (2.32 ms for AMD R7 4800U and 1.95 ms for NVIDIA RTX 4060), which is comparable to ThermTransformer and GPUThermalMap. Considering its substantial improvement in estimation accuracy while maintaining near real-time inference capability, TherMapNet achieves a superior trade-off between accuracy and efficiency, making it a promising solution for fast and accurate thermal estimation during runtime thermal management.

\begin{table*}[tbp]
\centering
\caption{Comparisons between TherMapNet and state-of-the-art NN-based thermal simulators for thermal estimations of the AMD R7 4800U CPU and the NVIDIA RTX 4060 GPU. Accuracy is measured by RMSE ($^\circ$C).}
\label{tab:rmse_comparison}
\begin{tabular}{l ccc c ccc c}
\toprule
\multirow{2}{*}{\textbf{Thermal Simulator}} 
& \multicolumn{4}{c}{\textbf{AMD R7 4800U}} 
& \multicolumn{4}{c}{\textbf{NVIDIA 4060}} \\
\cmidrule(lr){2-5} \cmidrule(lr){6-9}
 & Mean RMSE & Max RMSE & Std. of RMSE & Time (ms)
 & Mean RMSE & Max RMSE & Std. of RMSE & Time (ms) \\
\midrule
RealMaps~\cite{sadiqbatcha2021realmaps} & 2.190 & 19.762 & 2.648 & \textbf{0.87} & 2.075 & 9.885 & 1.892 & \textbf{0.65}\\
ThermGAN~\cite{jin2020ThermalGAN} & 0.596 & 10.880 & 0.985 & 4.25 & 0.397 & 2.323 & 0.252 & 4.13\\
ThermTransformer~\cite{lu2023ThermalTransformer} & 0.360 & 2.776 & 0.234 & 2.14 & / & / & / & /\\
GPUThermalMap~\cite{lu2025GPUThermalMap} & / & / & / & / & 0.190 & 0.812 & 0.132 & 1.89\\
Ours & \textbf{0.251} & \textbf{1.429} & \textbf{0.185} & 2.32 & \textbf{0.165} & \textbf{0.805} & \textbf{0.080} & 1.95\\
\bottomrule
\end{tabular}
\end{table*}

\subsection{Ablation Experiment}
The effectiveness of TherMapNet has been comprehensively demonstrated in the previous subsections through comparisons with IR measurements and state-of-the-art NN-based thermal simulators. To shed light on the contribution of key components in TherMapNet (i.e., positional encoding, input representation, and model architecture design), ablation experiments are conducted.

\subsubsection{Position Encoding}
\begin{figure}[tbp]
    \centering
    \includegraphics[width=1\linewidth]{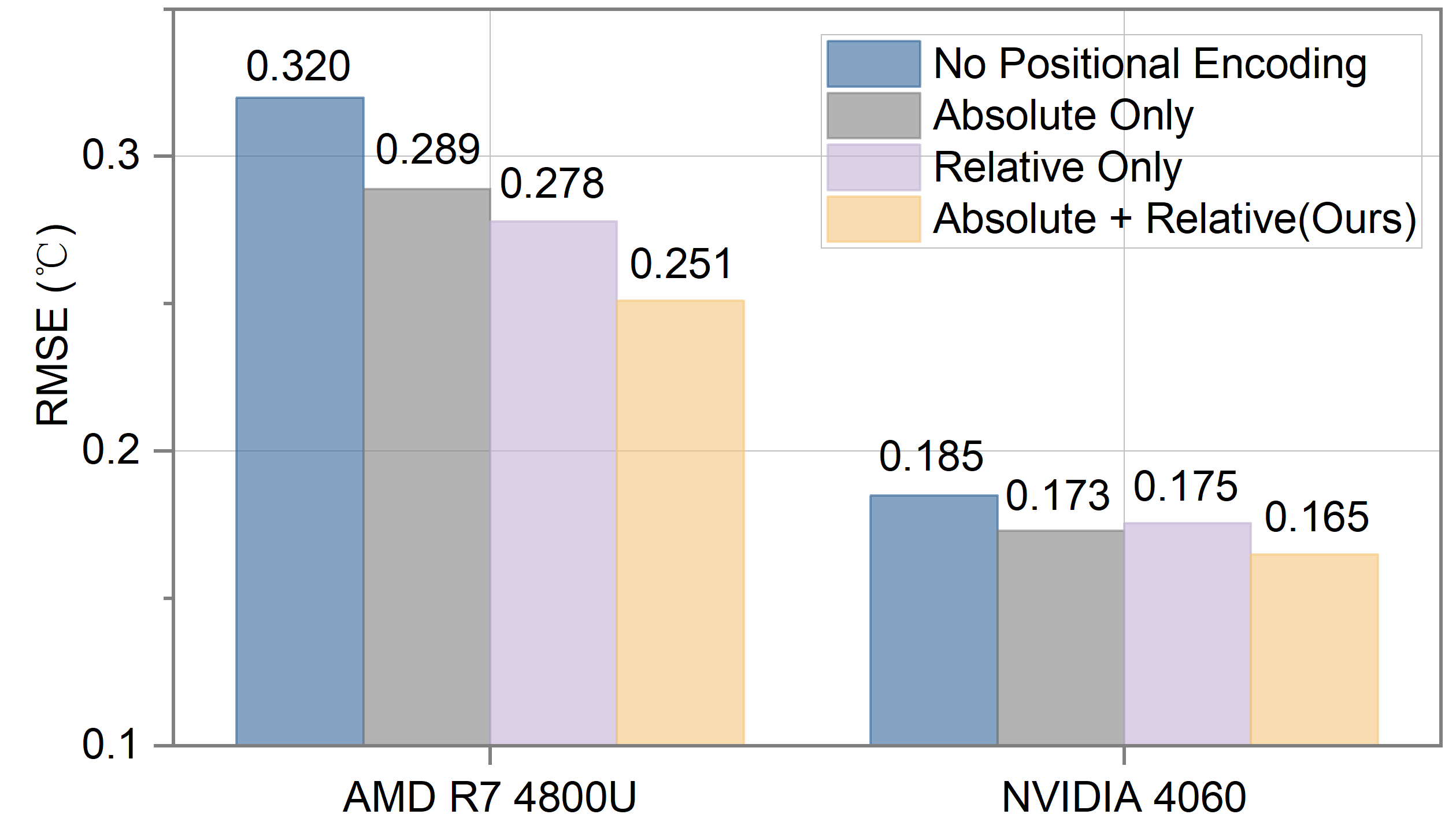}
    \caption{
    RMSE of ablation study on positional encodings.
    }
    \label{fig:pos_encoding}
\end{figure}

Fig.~\ref{fig:pos_encoding} presents the impact of different positional encoding strategies on the accuracy of TherMapNet. The results show that positional encoding is essential for accurate thermal estimation, as removing it leads to the largest RMSE on both datasets. Introducing either absolute or relative positional encoding significantly improves the estimation accuracy, demonstrating the importance of positional information. By combining both absolute and relative positional encoding, TherMapNet achieves the best performance, reducing the RMSE to 0.251$^\circ$C and 0.165$^\circ$C for the AMD R7 4800U and NVIDIA RTX 4060, respectively, which confirms the complementary benefits of the two positional encoding schemes.

\subsubsection{Token Representation}
To evaluate the effectiveness of the proposed token representation strategy, an ablation experiment is conducted using the conventional token representation, where all performance metrics at each time instant are treated as a token, and the results are compared with those of TherMapNet. As shown in Table~\ref{tab:token_ablation}, the conventional representation results in a degradation in accuracy, indicating its limited capability in capturing temporal dependencies. In contrast, TherMapNet treats the complete temporal sequence of each performance metric as a token, allowing the model to learn temporal correlations more effectively and thereby achieving superior thermal estimation accuracy.

\begin{table}[tbp]
\centering
\caption{Ablation study on token representation (RMSE in °C).}
\label{tab:token_ablation}
\begin{tabular}{lcc}
\hline
\textbf{Token Representation} & \textbf{AMD R7 4800U} & \textbf{NVIDIA 4060} \\
\hline
Original Input      & 0.295            & 0.173 \\
Transposed Input(Ours)    & \textbf{0.251}   & \textbf{0.165} \\
\hline
\end{tabular}
\end{table}

\subsubsection{Model Architecture}
The contribution of the CNN module and the DACM design is investigated and the ablation results are presented in Table~\ref{tab:cnn_branch}. It reveals that the Transformer-only model exhibits the largest RMSE, demonstrating its limited capability in capturing spatial characteristics of thermal maps. When a single CNN branch is integrated, the accuracy is significantly improved by enhancing local spatial feature extraction. The best accuracy is achieved by the proposed dual-branch CNN architecture, reducing the RMSE to 0.251$^\circ$C and 0.165$^\circ$C for the AMD R7 4800U and NVIDIA RTX 4060, respectively. This improvement verifies that the DACM design effectively captures complementary features.

\begin{table}[t]
\centering
\caption{Ablation study on CNN module (RMSE in °C).}
\label{tab:cnn_branch}
\begin{tabular}{lcc}
\hline
\textbf{Model Variant}        & \textbf{AMD R7 4800U} & \textbf{NVIDIA 4060} \\
\hline
Transformer only              & 0.320 & 0.184 \\
Single-branch CNN             & 0.273 & 0.171 \\
Dual-branch CNN (ours)        & \textbf{0.251} & \textbf{0.165} \\
\hline
\end{tabular}
\end{table}

\section{Conclusion}
In this work, TherMapNet, a Transformer-CNN cooperative thermal simulator, is proposed for real-time full-chip thermal map estimation directly from performance metrics. The Transformer employs a metric-centric tokenization strategy and positional encoding to effectively capture temporal dependencies and variations, while the CNN module enhanced by DACM and triplet loss focuses on extracting fine-grained spatial features. By integrating the complementary strengths of the Transformer and CNN modules, TherMapNet achieves accurate and efficient full-chip thermal map estimation. The effectiveness of TherMapNet is demonstrated by performing thermal estimations of the AMD R7 4800U CPU and NVIDIA RTX 4060 GPU, compared with IR measurements and state-of-the-art NN-based thermal simulators. Experimental results confirm that TherMapNet achieves high agreement with the measured thermal maps and outperforms existing NN-based methods in terms of accuracy while maintaining millisecond-level inference time. Furthermore, ablation studies verify the effectiveness of the proposed positional encoding strategy, metric-centric token representation, and Transformer-CNN cooperative architecture, demonstrating that each component contributes to the superior performance of TherMapNet.

\bibliographystyle{ieeetr}
\bibliography{Bibio}

\end{document}